\newcommand{\PaperNumber}{174}
\newcommand{\PaperTitle}{IPv6 Hitlists at Scale: Be Careful What You Wish For}
\documentclass[nonacm,sigconf,9pt]{acmart}
\usepackage{color}
\usepackage{graphicx}
\usepackage[labelformat=simple]{subcaption}
\usepackage{xspace}
\usepackage{multirow}
\usepackage{makecell} %
\usepackage{enumitem} %
\usepackage{multicol}
\usepackage[ruled,vlined]{algorithm2e}
\usepackage{cleveref}
\usepackage{pifont}%
\usepackage{ulem}
\usepackage[nolist,nohyperlinks]{acronym}
\newcommand{\etal}{et~al.\xspace}

\newcommand{\eg}{{e.g.,\@ }}

\newcommand{\parhead}[1]{\medskip \noindent \textbf{#1}\hskip .1in}

\newcommand{\vfour}{IPv4\xspace}
\newcommand{\vsix}{IPv6\xspace}
\newcommand{\icmpsix}{ICMPv6\xspace}
\newcommand{\eui}{EUI-64\xspace}
\newcommand{\ds}{dataset\xspace}
\newcommand{\hitlist}{IPv6 Hitlist\xspace}

\newcommand{\yarrp}{Yarrp\xspace}
\newcommand{\zmap}{ZMap\xspace}
\newcommand{\zmapsix}{ZMap6\xspace}
\newcommand{\maxmind}{MaxMind\xspace}

\newenvironment{widelist}{
	\begin{list}{$\bullet$} {
		\setlength{\leftmargin}{.40cm}
		\setlength{\itemsep}{.00cm}
	}}{\end{list}}

\ifdefined\isFinalized
\newcommand{\NewCommentType}[3]{\expandafter\newcommand\csname #1\endcsname[1]{{}}}
\else
\newcommand{\NewCommentType}[3]{\expandafter\newcommand\csname #1\endcsname[1]{{\color{#2}{#3: ##1}}}}
\fi

\usepackage[override]{cmtt} %

\copyrightyear{2023}
\acmYear{2023}
\setcopyright{rightsretained}
\acmConference[ACM SIGCOMM '23]{ACM SIGCOMM 2023 Conference}{September 10, 2023}{New York, NY, USA}
\acmBooktitle{ACM SIGCOMM 2023 Conference (ACM SIGCOMM '23), September 10, 2023, New York, NY, USA}
\acmDOI{10.1145/3603269.3604829}
\acmISBN{979-8-4007-0236-5/23/09}
\begin{document}

\newcommand{\name}{$\mathsf{SystemName}$\xspace} %
\newcommand{\Name}{$\mathsf{SystemName}$\xspace} %

\title{\PaperTitle}
\author{Erik Rye}
\affiliation{%
  \institution{University of Maryland}
  \country{}
  }

\author{Dave Levin}
\affiliation{%
  \institution{University of Maryland}
  \country{}
  }

\begin{abstract}
Today's network measurements rely heavily on Internet-wide scanning, employing tools
like ZMap that are capable of quickly iterating over the entire IPv4 address space.
Unfortunately, IPv6's vast address space poses an existential threat for
Internet-wide scans and traditional network measurement techniques.
To address this reality, efforts are underway to develop ``hitlists'' of
known-active IPv6 addresses to reduce the search space for would-be scanners.
As a result, there is an inexorable push for constructing as large and complete a
hitlist as possible.

This paper asks: what are the potential benefits and harms when IPv6
hitlists grow larger?
To answer this question, we obtain the largest IPv6 active-address list to
date: 7.9 billion addresses, 898 times larger than the current
state-of-the-art hitlist.
Although our list is not comprehensive, it is a significant step forward and
provides a glimpse into the type of analyses possible with more complete
hitlists.

We compare our dataset to prior IPv6 hitlists and show both benefits and
dangers.
The benefits include improved insight into client devices (prior datasets
consist primarily of routers), outage detection, IPv6 roll-out, previously
unknown aliased networks, and address assignment strategies.
The dangers, unfortunately, are severe: we expose widespread instances
of addresses that permit user tracking and device geolocation, %
and a dearth of firewalls in home networks.
We discuss ethics and security guidelines to ensure a safe path towards more
complete hitlists.

 \end{abstract}



\maketitle

\thispagestyle{empty}
\pagestyle{empty}
\section{Introduction} 
\label{sec:intro}

\zmap~\cite{durumeric2013zmap} revolutionized Internet scanning in
2013, enabling scans of the entire \vfour Internet in under an hour.
Since then, Internet-wide scanning has become one of the most powerful
and commonly used tools for measurement researchers and practitioners,
leading to previously inaccessible findings in
security~\cite{dns-manipulation,weaponizing,durumeric2014matter,adrian2015imperfect,costin2014large}, topology
discovery~\cite{imc16yarrp}, IoT measurement~\cite{detecting-iot},
outage detection~\cite{timeouts}, and more.

Unfortunately, the continued, accelerating deployment of IPv6 represents an existential crisis
for Internet-wide
scanning.
In contrast to IPv4, brute-force scanning of every IPv6 address is
impossible owing to 
IPv6's exponentially larger address space ($2^{128}$ compared to \vfour's $2^{32}$).

Without the prospect of iterating over all IPv6 addresses, some scanning
tools instead rely on ``hitlists'' that identify addresses that are
likely to be active and in-use, and probe only those. Others have
introduced \vsix target generation algorithms that emit
potentially-active \vsix addresses as probing candidates; these models must be trained on
\emph{some} hitlist and are biased to the types of addresses contained in their
training data.
Thus, the larger and more representative these hitlists are, the more complete the view of the IPv6 Internet
available to measurement efforts~\cite{richter-v6-scanning}.

As a result, the ``holy grail'' of Internet scanning is a complete list
of all active IPv6 addresses. There are ongoing efforts approximate 
such a list.
To date, the largest public list is the \emph{IPv6
Hitlist}~\cite{expanse,rusty}. The IPv6 Hitlist uses a variety
of active measurement techniques---namely \zmapsix~\cite{zmap6} and
\yarrp~\cite{imc16yarrp}---and target generation algorithms to
discover new responsive IPv6 addresses.\footnote{We use capitalization to
differentiate between hitlists in general (lowercase) and the IPv6
Hitlist in particular (uppercase).}
Their most recent efforts in 2022 nearly tripled IPv6 Hitlist's size to a total of
8.8M addresses~\cite{rusty}.

In this paper, we ask: what are the potential benefits and harms when
IPv6 hitlists grow larger?
To answer this question, we obtain the largest list of active IPv6
addresses to date: 7.9 billion addresses, 898 times larger than the
current IPv6 Hitlist snapshot and 370 times larger than the Hitlist
over the same time interval.
Our corpus contains nearly 15 times more active \vsix addresses than \emph{all} of the active IPv6
addresses discovered in the IPv6 Hitlist's four-year history though it
was collected in about 15\% of the time.
In contrast with the active measurements employed by the IPv6 Hitlist, we
passively collected the nearly 8 billion active IPv6 addresses in our corpus by
running a set of 27 geographically-distributed NTP servers as part of the NTP
Pool~\cite{ntppool}, which devices worldwide use to synchronize their clocks. 

Although our hitlist is not comprehensive---for instance, we are likely
missing most modern Android devices because they are not configured to use
the NTP Pool by default---it is a significant leap forward from the current
state of IPv6 hitlists.
As such, it provides a glimpse into what the future holds as
the community moves toward its goal of a more comprehensive hitlist.

Our glimpse into the future of IPv6 hitlists produces two broad
results: that network measurement insights are indeed improved through
a bigger list (especially one that is passively collected like ours),
and that large hitlists poses significant security and privacy
risks.

\parhead{Benefits: New insights from larger hitlists} %
What is gained from having larger IPv6 hitlists?
What do current hitlists lack, and does filling in those gaps facilitate 
a deeper understanding of the IPv6 Internet?

We find that a significant missing portion of prior IPv6 hitlists is
\emph{end-host addresses}.
The IPv6 Hitlist, for example, relies primarily on
\zmapsix and \yarrp, and as a result discovers many infrastructure
nodes (especially routers and CPE), but has difficulty identifying end-hosts due
to firewalling and frequently unpredictable and ephemeral client
addresses~\cite{rfc3041}.
Conversely, our dataset is comprised largely of end-hosts, though not
exclusively (virtually all Internet devices must synchronize their
clocks).

We demonstrate that when a hitlist has more end-host addresses, it
enables investigations that are impossible with existing
hitlists, such as studying client address entropy, IoT detection, and address
assignment patterns.
However, these new insights are not without cost.

\parhead{Threats: New privacy leaks from larger hitlists} %
Conventional wisdom dictates that IPv4 addresses do not constitute
personally identifiable information (PII).
This is largely because the link between a device and an IPv4 address
is very weak: addresses are assigned randomly and are frequently reassigned, and
multiple devices often map to the same address at any time.
However, the same is not always true for IPv6 addresses.
It is now well-known that IPv6 addresses can risk uniquely identifying
a client device---e.g., if the device embeds its MAC address into the
lower-order bits of its IPv6 addresses, then it could potentially be
tracked as it moves across networks.

These privacy issues were not a major concern for previous IPv6
hitlists, as they focused primarily on infrastructure nodes, which
typically have no active user and do not move much across networks.
However, more complete hitlists will inevitably include more client
devices.

We perform what we believe to be the first analysis of the privacy
risks inherent to large, client-rich IPv6 hitlists. We observe nearly 15 million
clients in multiple networks across four distinct types of address tracking, and
apply recent precision \vsix geolocation techniques to hundreds of thousands of
addresses.

\medskip \noindent
Collectively, our results show that there is significant promise in
store for larger IPv6 hitlists and Internet scanning, but also
potential harm in making larger hitlists public.

\parhead{Contributions} %
To summarize, we make the following contributions:

\begin{widelist}
\item We collect and report on the largest publicly-obtained IPv6
	hitlist to date: over three orders of magnitude larger than prior
	public hitlists.

\item We perform a thorough comparison with the two premiere hitlists in
	use today---the \emph{IPv6 Hitlist}~\cite{rusty} and CAIDA's routed
  /48 dataset~\cite{caida-routed48}---finding that our NTP-derived dataset is
  far larger and comprises more end-hosts, but that each dataset provides a
  complementary perspective on active IPv6 addresses.

\item We show that larger, more client-centric IPv6 hitlists enable new
	discoveries that prior hitlists do not, particularly in address
	patterns and aliased network discovery.

\item We also show that larger, more client-centric IPv6 hitlists
	enable new \emph{attacks} on privacy in the form of tracking and
	geolocation.

\item We discuss the ethical ramifications of our findings, and provide
	guidelines that we hope will help shape future efforts in obtaining
	and sharing more complete IPv6 hitlists.

\item We make the active /48 prefixes we discovered publicly available
	at \url{https://v6-research.cs.umd.edu}
	
\end{widelist}


\section{Background and Related Work} 
\label{sec:background}

\subsection{Why creating IPv6 hitlists is hard}

Applications ranging from census and adoption
studies~\cite{dynamic,Czyz-adoption}, to vulnerability
identification and remediation~\cite{markowsky2015scanning}, to outage
detection~\cite{schulman2011pingin,richter2018advancing,failures,Luckie:2017:IRO:3098822.3098858}
all rely on an understanding of what IP addresses are assigned to live hosts. In
\vfour, identifying these hosts is tractable due to stateless scanners like
\zmap~\cite{durumeric2013zmap} and \yarrp~\cite{imc16yarrp} that can probe the
\vfour Internet in minutes. However, several factors complicate live address
discovery and hitlist creation in \vsix. 

The first and most obvious complicating factor is the immensity of the \vsix
address space. This leads to large prefixes---at least /64 and frequently /56
or larger~\cite{bcop-prefix}---being delegated to even residential customers.
Thus, the average Internet subscriber has 4 billion or more times the number of
IP addresses in the entire IPv4 Internet available to her, all of which are
publicly routable; contrast this with \vfour, in which she commonly has one
public IP, with her home network hidden behind a NAT. With such large prefixes
delegated to customers with only tens or hundreds of devices, \vsix is
considerably more 
sparse than \vfour. Further, some service providers delegate prefixes to their
customers for only short periods (\eg 24 hours) before recuperating them and issuing new
ones~\cite{rye2021follow}.

The lower 64 bits of a 128-bit \vsix address is called the Interface Identifier
(IID), and its assignment adds further complexity to creating \vsix
hitlists. Some \vsix addresses are manually assigned: these are often
infrastructure devices that network administrators prefer to assign easily
memorable IIDs for troubleshooting~\cite{v6exhaust-ic16}, like
\texttt{::1} or \texttt{::2}. Occasionally, some network operators will embed
the \vfour address assigned to the same interface in the
IID~\cite{siblings-pam15}, but there is no requirement to do so and these
embeddings are relatively uncommon. Often, hosts self-assign IIDs via one
of several processes. Extended Unique Identifier - 64 (EUI-64) Stateless Address
Autoconfiguration (SLAAC)~\cite{rfc2462} embeds the interface's Media Access
Control (MAC) address in the IID, after inverting the Universal/Local bit of the
MAC and inserting a \texttt{0xFF} \texttt{0xFE} in-between the third and fourth bytes.
Because these addresses embed a static, link-layer identifier in the \vsix
address that allows for device identification and tracking, as well as attacks
tailored to device manufacturers, generating ephemeral random addresses has
instead been encouraged since 2001~\cite{rfc3041}. However, ephemeral addresses
are problematic for servers, which should possess stable yet
privacy-preserving addresses for long periods of time. In response, standards
have been proposed for generating stable, random addresses~\cite{rfc7217}.
Finally, DHCPv6~\cite{rfc8415}, the less-ubiquitous cousin of its \vfour
analog, also exists to assign addresses to hosts. However, given such large
prefix allocations, it is up to the DHCPv6 implementation and operator
configuration how client addresses are assigned.  

Even responsive addresses in \vsix may not indicate a live host.
In \vsix, \emph{aliasing}, in which a single device replies to probes to an
entire network, is a relatively common practice. This necessitates a filtering
step in hitlist creation, wherein responsive addresses from aliased networks
are removed.

\subsection{Related Work}
\label{sec:related}

There have been extensive efforts to discover new live IPv6 addresses through both
active and passive measurements.

\parhead{Active approaches}
Beverly developed the \yarrp high-speed, stateless traceroute tool to improve
host and topology discovery in both \vfour and \vsix~\cite{imc16yarrp}, and
Beverly \etal used \yarrp to discover significant \vsix Internet core
topology~\cite{imc18beholder}. Similarly, Gasser \etal developed \zmapsix~\cite{zmap6} \vsix
extensions of the \zmap~\cite{durumeric2013zmap} high-speed scanner to enable
fast \vsix scanning without tracing to intermediate hops~\cite{gasserscanning}.
Gasser \etal used a combination of \zmapsix, scamper
traceroutes~\cite{luckie2010scamper}, and public data sources to develop an
\vsix hitlist and identify aliased
networks~\cite{expanse}. They continue to publish a
weekly hitlist of responsive addresses and known aliased and non-aliased
networks~\cite{hitlistwebsite}. Rye and Beverly used \yarrp and \zmapsix to
focus discovery of topology at the network periphery (\ac{CPE} devices in
customer ISP networks)~\cite{edgy} and characterized high-frequency customer
network changes~\cite{rye2021follow}. Others have enumerated reverse DNS zones
to discover active \vsix
addresses~\cite{pam2017-nxdomain,BorgolteSP2018,strowes2017bootstrapping}.
Foremski \etal~\cite{Foremski:2016:EUS:2987443.2987445} aggregated datasets
comprising more than 3.5 billion \vsix addresses from cloud providers and ISPs
to develop new candidate addresses for active measurements. %
Numerous machine learning models~\cite{steger2023target} have been trained on
responsive addresses in order to generate candidate addresses for active
measurements, using Bayesian networks~\cite{6gen}, reinforcement
learning~\cite{6hit}, generative adversarial networks~\cite{6gan}, divisive
hierarchical clustering~\cite{6tree}, and ensemble learning~\cite{6forest}.

As we will demonstrate, our passive approach is largely complementary to these
active efforts, exposing portions of the in-use \vsix address space that the
active efforts alone do not reach.

\parhead{Passive approaches}
Numerous passive \vsix efforts also have studied
\vsix addressing. 
Gasser~\etal~\cite{expanse} and
Huz~\etal~\cite{huz2015experience}
both crowd-source small
numbers of \vsix client addresses via Mechanical
Turk~\cite{mturk} and
Prolific Academic~\cite{proa}.
A major barrier to conducting large-scale passive \vsix measurements, however, is access to
proprietary datasets. Plonka and Berger~\cite{Plonka:2015:TSC:2815675.2815678} use \vsix addresses
gathered from a large CDN's webservers to determine how customer addresses
change over time.  Using data obtained from Facebook, Li and
Freeman~\cite{imc20facebook} examine how client \vsix addresses change over time
and consider the problem of handling abusive or malicious actors in \vsix. %
Saidi \etal~\cite{gasserapple} examine aggregated \vsix client traffic,
including NTP, provided by a large European ISP to track customer subnet
allocations over time by tracking clients employing \eui \vsix addresses. Enayet
and Heidemann use data from the DNS B root to detect outages in \vfour and
\vsix~\cite{enayetoutages}, while Fukuda and Heidemann use DNS backscatter from
the B root to detect \vsix scanning~\cite{fukudascanning}. 

\parhead{Comparison of passive and active approaches}
Several works compare active and passive measurement approaches, albeit only in
IPv4 and on a far smaller scale than our work. Bartlett~\etal compared passive
monitoring and active probing to discover services running on a university
network~\cite{bartlettunderstanding}; Heidemann~\etal similarly used passive and active approaches to
discover IPv4 end hosts in 2008~\cite{heidemann2008census}. Zander~\etal used
several passive sources of active \vfour addresses, including Wikipedia edits and
NetFlow records, to augment active measurements to estimate IPv4 address space
utilization~\cite{zander2014capturing}.

Unlike prior active efforts, we avoid sending millions of unsolicited probes in
search of active addresses. Unlike many prior passive efforts, our technique does not require access to
privileged datasets obtained by private organizations.
Rather, we demonstrate that enormous amounts of \vsix data can be obtained by
contributing to an open service---the NTP Pool---which we review next.

\subsection{NTP and the NTP Pool} %

The \acf{NTP} is one of the Internet's oldest protocols, standardized in 1985
in RFC958~\cite{RFC958}. NTP synchronizes a
device's clock with a remote time server, even in variable-delay networks.
Keeping accurate time is of immense importance to a variety of applications
ranging from TLS certificate
validation~\cite{zhang2014analysis,malhotra2015attacking}, to
authentication~\cite{malhotra2015attacking,morowczynski2012did,kohl1993kerberos}, to DNS cache
entries~\cite{malhotra2015attacking}. Most devices on the Internet
today synchronize their clocks using NTP.

Where
a device looks for its time is typically a function of its operating system. Windows
clients and servers, for instance, synchronize their time with
\texttt{time.windows.com}~\cite{windowstime} by default when not joined to a domain.
Likewise, Apple devices synchronize with \texttt{time.apple.com}.
Android clients until Android 8 (Oreo)
used the NTP Pool (\texttt{pool.ntp.org}, discussed next); later versions now use
\texttt{time.android.com}~\cite{androidtime}. NTP
servers can additionally be specified via DHCP~\cite{rfc2132}
and DHCPv6~\cite{rfc5908} options.

The NTP Pool Project~\cite{ntppool} provides NTP service through a worldwide set
of geographically distributed NTP servers, many of which are contributed by
volunteers. Any host with a publicly reachable IP address can serve in the NTP
Pool.  The Pool preferentially directs clients to servers geographically near
them, using a combination of IP geolocation and DNS round robin. 
The NTP Pool Project further provides various ``vendor zones'' to equipment and
software vendors to use as defaults on their devices; vendor zones for Android,
Ubuntu, and CentOS exist, among many others (\texttt{android}, \texttt{ubuntu},
and \texttt{centos}\texttt{.pool.ntp.org}, respectively.)

%
 %


\section{Methodology} 
\label{sec:methodology}

\begin{table*}[t]
	\small
\begin{tabular}{lcrrrrrrr}
\multicolumn{1}{c}{\multirow{2}{*}{\textbf{Dataset}} } & \multirow{2}{*}{\textbf{Dates}} & \multicolumn{2}{c}{\textbf{IPv6 Addresses}} & \multicolumn{2}{c}{\textbf{ASNs}} & \multicolumn{2}{c}{\textbf{/48s}} & \multicolumn{1}{c}{\multirow{2}{*}{\textbf{\begin{tabular}[c]{@{}c@{}}Avg. Addrs\\ per /48\end{tabular}} }} \\
\multicolumn{1}{c}{} &  & \multicolumn{1}{c}{Num.} & \multicolumn{1}{c}{Common} & \multicolumn{1}{c}{Num.} & \multicolumn{1}{c}{Common} & \multicolumn{1}{c}{Num.} & \multicolumn{1}{c}{Common} & \multicolumn{1}{c}{} \\
\hline
NTP Pool (This paper) & Jan–Aug '22 & \textbf{7,914,066,999} & -- & 9,006 & -- & 7,205,127 & -- & \textbf{1,098} \\
    IPv6 Hitlist~\cite{hitlistwebsite} & Feb–Aug '22 & 21,409,629 & 277,026 & \textbf{18,184} & 7,560 & 431,851 & 267,908 & 50 \\
    CAIDA Routed /48~\cite{caida-routed48} & Feb–Apr '22 & 11,613,494 & 3,117 & 13,770 & 6,957 & 11,111,563 & 102,864 & 1
\end{tabular}

	\vspace{.1in}
	\caption{Comparison \vsix datasets considered. Our NTP corpus
	is passively collected, while the comparison datasets used active techniques.
	``Common'' denotes the intersection of the comparison data with our data.}
    \label{tab:datasets}
\end{table*}

This section first highlights the measurement infrastructure we established to conduct
our experiments. Then, it introduces the other \vsix address datasets that we
compare to our NTP-derived corpus. Finally, we discuss how we geolocate NTP
client addresses and our methodology for probing back to active NTP clients that
query our servers.

\parhead{Vantage Points}
In order to measure the effectiveness of using NTP servers as
large-scale, longitudinal, passive \vsix measurement infrastructure, we
operated 27 NTP servers from 25 January 2022 through 31 August 2022.  
We chose \acp{VPS} from 20 countries across 6 continents to obtain
geographic diversity.
Specifically, we ran 6 servers in the US, 2 in Japan, 2 in Germany, and 1 server in
each of: Australia, Bahrain, Brazil, Bulgaria, Hong Kong, India,
Indonesia, Mexico, Netherlands, Poland, Singapore, South Africa, South
Korea, Spain, Sweden, Taiwan, and the United Kingdom.

Though we ran servers in 20 countries, the NTP Pool's load balancing allowed us to
collect data from 175 countries\footnote{We count ISO-3166-1 two-letter country
codes and use the term ``countries,'' although some are dependent territories.} in total.
The majority of the IPv6 addresses we discovered came from India
(1.9B), China (1.6B), US (1.2B), Brazil (700M), and Indonesia (630M),
collectively accounting for 76\% of our entire dataset.
The other 170 countries accounted for 24\%.

We emphasize that each \ac{VPS} was minimally provisioned and cost on average
\emph{less than \$7 per month} to operate.
We typically used one virtual core, 500MB--2GB of
RAM, and a Linux OS available from the VPS vendor (Ubuntu, Amazon
Linux, or CentOS).

Each of these \acp{VPS} was configured as a stratum-2 NTP server and joined to
the NTP Pool. Because the NTP Pool directs clients to its NTP servers via a DNS
round-robin that incorporates the client's coarse-grained IP
geolocation, our globally-distributed NTP servers were visited by a wide range
of clients around the world.

\parhead{Comparative Datasets}
In order to compare our passive NTP results with contemporaneous,
state-of-the-art \vsix measurements, we acquired two external datasets.
First, we compare against the IPv6 Hitlist~\cite{expanse},
which provides a list of responsive addresses and networks that the operators
detect as being aliased networks (responsive on all addresses) and
those that are not.
The Hitlist is updated on roughly a weekly basis, so in order to best compare our \ds to
their data, we consider all Hitlist responsive \vsix addresses published during
our study's time frame. \hitlist data concurrent with our study was published first on 16 February 2022 and
runs through 29 August 2022. While our study collects only NTP requests, the
\hitlist is obtained through active measurements using ICMPv6, TCP ports 80 and
443 (HTTP and HTTPS), and UDP ports 53 (DNS), 161 (SNMP), and 443 (QUIC). Our
\hitlist comparison \ds consists of 21,409,629 unique \vsix addresses.

Second, we leverage a large dataset of 1,083,188,032 \yarrp traces conducted
by CAIDA from their Archipelago distributed measurement system~\cite{caida-ark}
between 3 February and 6 April 2022. %
Their measurement methodology splits each prefix of length /32 or longer into /48s and
probes the \texttt{::1} address of each /48. For prefixes of length less than /32, only a
single \texttt{::1} address is probed, with no splitting into constituent /48s.
These traces discovered 11,613,494 live
addresses. We refer to this measurement as the ``CAIDA routed /48'' dataset
throughout.

Table~\ref{tab:datasets} lists the relevant details of each \ds involved in our
study.
We will explore these numbers in more detail in
\S\ref{sec:measurements}, but even at a glance it is clear that the NTP
data corpus comprises three orders of magnitude more addresses, with
a greater density of addresses on average in each /48.

\parhead{Geolocation}
We consider two different types of geolocation in our results. First, we used
\maxmind's GeoLite2 City database~\cite{maxmind} to geolocate the NTP client
addresses we observed.
While fine-grained IP-geolocation is often error-prone, particularly in \vsix,
we consider only the country reported by \maxmind in aggregated results and do not use
the more granular geolocation data it reports.

Second, for \vsix addresses with embedded MAC addresses in the form of \eui
\acp{IID}, we attempt to link this embedded MAC address with a wireless MAC
address from the same device obtained from a geolocation service. \eui \acp{IID}
are constructed by first inverting the seventh least significant bit of the most
significant byte of the interface's MAC address. Then, a static \texttt{OxFF
0xFE} is inserted between the third and fourth bytes of the MAC address to
create a 64-bit identifier; this is then used as the lower 64 bits of the IP
address in an \eui \vsix address. Recovering interface MAC addresses is a simple
process --- the \texttt{0xFF 0xFE} bytes of an \eui \acp{IID} are removed,
followed by the seventh bit's inversion.

Both Google and Apple both offer geolocation APIs~\cite{googlegeo,applegeo}, and
other individual and community projects also collect wireless geolocation 
information~\cite{wigle,openwifi,mylnikov,openbmap}. Our methodology for linking
wired \eui-derived MAC addresses to wireless MAC addresses follows that
of Rye and Beverly~\cite{ipvseeyou}, wherein they form a linkage between the
most commonly-arising offsets between pairs of wired and wireless identifiers
from within the same vendor-assigned three-byte address prefix, called an
Organizationally Unique Identifier (OUI). This is often, but not always, the closest match
between wired and wireless MAC addresses within the same OUI. This methodology
is also limited to devices that have wireless and wired MAC addresses from the
same OUI.

\parhead{Backscanning}
In order to compare and contrast active and passive methodologies for compiling
\vsix hitlists, we actively probed NTP clients that visited five of our 27 NTP
servers over the course of a week during January 2023. During this week, we
recorded the source addresses of NTP clients that queried the servers over ten
minute intervals. When the interval concluded, we initiated traceroutes from the
servers back to the clients using \yarrp and sent \icmpsix Echo Requests to the
clients with
\zmapsix. The probe targets were both the NTP client address that had queried
the NTP server, as well as a random \vsix address within the same /64 as the
client. All probes used \icmpsix in order to minimize potential disruption to
the probed addresses; no IP was probed more than once during a 10 minute
interval.

\parhead{Ethical Considerations}
\label{sec:ethics}
During this study, we accumulated nearly 8 billion unique \vsix addresses by
adding 27 NTP servers to the NTP Pool. Users of the NTP Pool had no way
of knowing their NTP request data could or would be used in our study.
That said, our study follows the same general principles of prior IPv6 Hitlist
generation~\cite{expanse} as well as other peer-offered
infrastructure, such as studies that use BitTorrent to collect and study IP
addresses~\cite{richter-cgnat}.
Like those studies, we do not collect any PII that might be included in the
application-layer data (NTP requests do not contain PII).

Novel to our findings, however, is that large sets of IPv6 addresses may in and
of themselves contain enough information to track and geolocate users.
These attacks on privacy are made possible through the lower-order bits
(specifically, we make use of \eui IIDs).
Thus, to avoid spreading this potentially sensitive information, we will only
be releasing our dataset at the /48 level.
This is an ethical consideration that future IPv6 hitlists must contend with:
what is an appropriate way to share hitlists so as to enable Internet scanning
tools to use them?

We communicated with the project owners of the NTP Pool to inform them of our
experiments and to ensure that we were abiding by both the terms of service and
acting in a way that preserved the privacy of the NTP Pool users.  They
concurred that we were not violating any NTP Pool policy or community standard
and requested that address data released be aggregated to protect user privacy.

We also submitted our study to our institution's IRB for review.
Much like other institutions' IRBs~\cite{expanse}, they
did not consider IP addresses as constituting human data.
However, after informing them of our privacy results (\S\ref{sec:security}),
they agreed that further future consideration would be appropriate.
Our hope is that this paper can be a first step towards the networking
community helping to guide appropriate methods for ethically sharing IPv6
hitlists as they continue to grow.

Finally, contrary to active measurements that introduce immense volumes of
superfluous data for the sole purpose of eliciting responses from remote
devices (\eg \texttt{traceroute} and \texttt{ping}), our experiments actually
\emph{provided a beneficial service} to the devices we measured. All of our
servers provided stratum-2 NTP service and are located in cloud providers with
exceptionally high availability, providing a reliable source of accurate time
for NTP clients. Therefore, we believe that the benefits of our work outweigh
the potential harm or risk that it may present.
%
%
%
%

%
%
%
%
%
%
%
%
%
%

%
%
%
%
%
%
%
%
%
 %
%


\section{Benefits of Larger Hitlists} 
\label{sec:measurements}

We begin our analysis by evaluating whether larger IPv6 hitlists confer
benefits: as the community races to obtain larger hitlists, is it worth
it?
To this end, we first show that our NTP-based dataset is not only
larger, but nearly disjoint with and complementary to other
state-of-the-art IPv6 measurements.
Then, we evaluate various applications of this larger hitlist:
measuring aliased networks and analyzing IPv6 addressing patterns.

\subsection{How Do the Datasets Compare?} %

First, we compare our \ds to the \hitlist and a large-scale active measurement
conducted by CAIDA.

\parhead{In terms of size}
Table~\ref{tab:datasets} compares the aggregate number of live \vsix
addresses we observed during our study to the number of responsive
addresses collected by the \hitlist project and during a large-scale,
active measurement campaign by CAIDA. 
Running 27 \vsix NTP servers from January through August 2022, we
observed 7.9 billion unique \vsix source addresses---370 times more
than the IPv6 Hitlist's collections over a comparable period of time. The CAIDA
measurement discovered 11,613,494 addresses during its two-month run, 681 times
fewer than the NTP corpus.

\parhead{In terms of addresses discovered} %
Despite the massive difference in size, our dataset did not  subsume either of the active datasets;
we only discovered 1.3\% (277,026) of the addresses that IPv6 Hitlist
found, and a mere 0.02\% of the IPv6 addresses CAIDA's routed /48 dataset discovered.
This shows that the datasets are indeed complementary, and suggests
that the kinds of devices we are finding are in fact distinct. We
confirm this hypothesis later in this section.

\parhead{In terms of \acp{AS}} %
While the number of raw addresses in the NTP corpus dwarfs the other
dataset address counts by several orders of magnitude, this trend is
reversed in the number of \acp{AS} we observe.
Our NTP corpus contains 65.3\% of the number of ASes observed in the
CAIDA scan (9,006 vs 13,770) and 49.5\% of the number discovered in the
\hitlist data (9,006 vs 18,184).
This discrepancy is likely due to the nature of the two active
datasets, which use \texttt{traceroute}-like tools to discover Internet
infrastructure between their vantage point and their probe targets.
Our data, coming from NTP clients, is concentrated in \acp{AS} where
NTP Pool clients exist, typically in customer ISPs.
This hypothesis is strengthened by examining the types of \acp{AS} the
different corpora addresses originate from, as classified by
ASdb~\cite{ziv2021asdb}.
While the top AS type is consistent between all three datasets
(``Computer and Information Technology'',``Internet Service Provider
(ISP)''), an additional 14\% (1,146,709,677) of our NTP Pool corpus
originates from ``Phone Provider'' ISP subtype.
By contrast, only 2\% of the \hitlist addresses originate from ``Phone
Provider'' \acp{AS}, indicating that the NTP Pool corpus consists of a
higher percentage of mobile clients than do either of the two active
datasets.

\parhead{In terms of prefixes} %
Though the NTP Pool corpus \vsix addresses are concentrated in fewer
\acp{AS} than either of the two active measurements, our NTP dataset
exhibits the highest number of addresses discovered per /48
(Table~\ref{tab:datasets}).
The NTP dataset discovers a mean of 1,098 addresses per /48, while the
\hitlist uncovers 50 and the CAIDA scanning only 1.
This ``address density'' has at least two potential root causes.
First, NTP Pool clients may more commonly be client devices that
frequently change their (random) \ac{IID} in order to prevent tracking, as is
considered best practice for client devices~\cite{rfc3041}.
This phenomenon would manifest as many different addresses originating
from the same prefix---for instance, a /56 or /64 allocated to a
residential deployment by a customer ISP.
A second possibility is that our passive NTP methodology detects more
customer deployments than either of the active probing methods.
While we detect devices in customer deployments so long as they visit
an NTP Pool server, these networks are often highly subnetted and would
require significant active probing to discover the CPE devices, which
may or may not respond.
Neither the \hitlist nor CAIDA scanning is specifically calibrated to
do such fine-grained, residential deployment discovery.
Note that these two root causes are not mutually exclusive; both may
manifest in our data.

\parhead{In terms of device type} %
It is in general very difficult to perform device fingerprinting at
scale.
However, we can fortunately make use of features of the IPv6 addresses
themselves to gain some insight into the kinds of devices that comprise
the respective datasets.

We examined the \acp{IID}, or lower 64 bits, of the addresses in each
\ds.
Because \acp{IID} uniquely identify a host interface on a network, and because
different types of hosts are subject to different concerns (\eg
preventing client tracking, static server addresses, or infrastructure
addresses easily memorable by network administrators), the randomness
(or lack thereof) of an \ac{IID} may vary significantly between device
types.

\begin{figure}[t]
\centering
    \includegraphics[width=\linewidth]{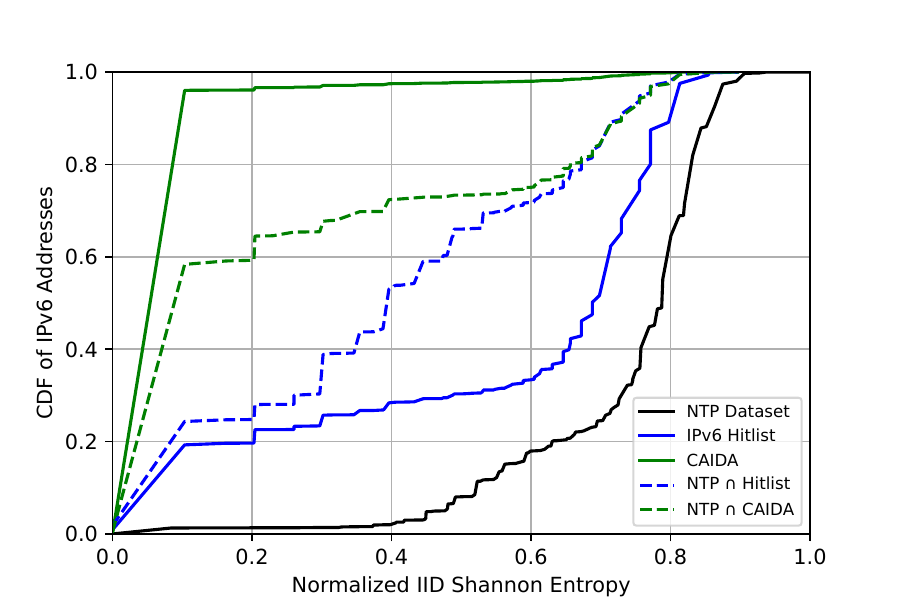}
    \caption{IID entropies of the \vsix addresses from the NTP Pool, \hitlist,
    and CAIDA routed /48 corpora, as well as the IID entropies of their
    intersections.}
\label{fig:entropy-compare}
\end{figure}
\begin{figure*}[t]
\centering
    \begin{subfigure}[c]{.48\textwidth}
        \includegraphics[width=\linewidth]{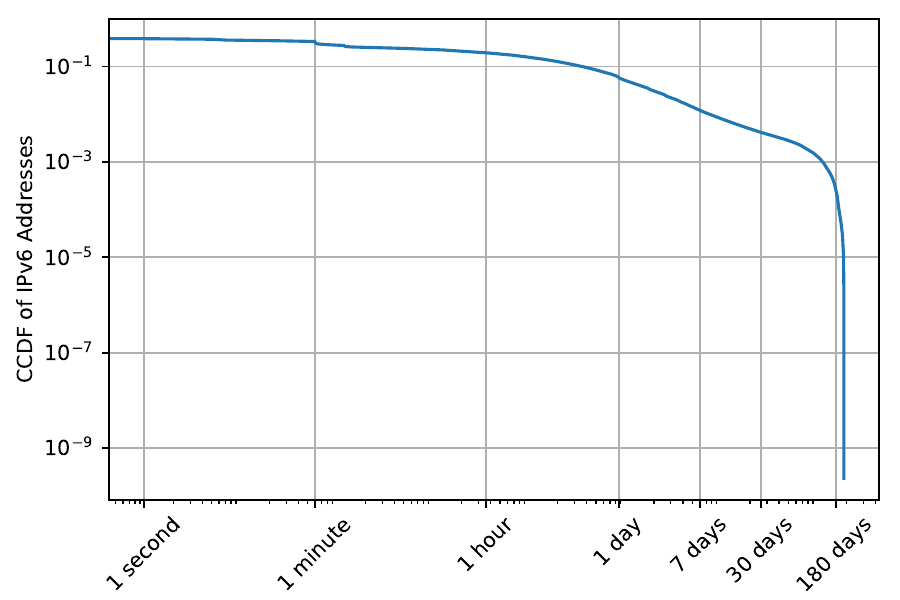}
        \caption{CCDF of distinct address lifetimes over all \vsix addresses observed
        during our study.}
        \label{fig:all-lifetimes}
    \end{subfigure}%
	\hfill
    \begin{subfigure}[c]{.48\textwidth}
        \includegraphics[width=\linewidth]{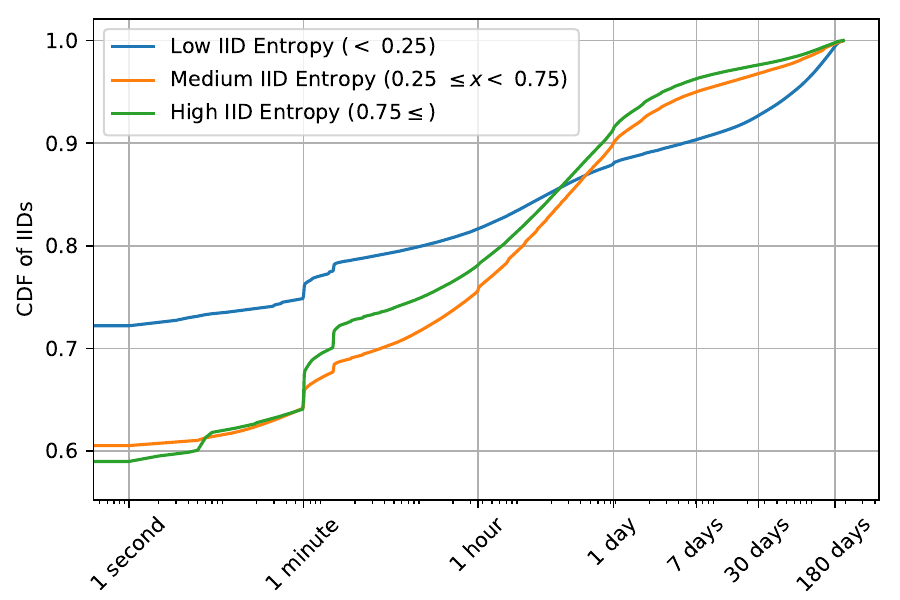}
        \caption{CDF of IID lifetimes by IID entropy category. }
        \label{fig:iid-lifetimes}
    \end{subfigure}%
    \caption{Address and IID lifetimes vary significantly. While most addresses
    and IIDs are observed only once, many persist for days or months.}
    \label{fig:lifetimes}
\end{figure*}

Figure~\ref{fig:entropy-compare} plots the CDF of all addresses found
in each dataset versus \ac{IID} entropy, using the normalized Shannon
entropy as a metric.
The NTP Pool corpus exhibits significantly higher entropy than the
other two datasets, with a median normalized Shannon entropy of
approximately 0.8.
The \hitlist has a somewhat lower median entropy of about 0.7, while
almost the entirety of the CAIDA dataset has extremely low entropy.

These data points reinforce the hypothesis that \emph{our NTP Pool
dataset consists primarily of client devices}, which often use
ephemeral, random addresses to defend against long-term tracking.
The CAIDA dataset, by contrast, discovers mainly core Internet
infrastructure, which is often manually addressed by operators with an
incentive to create easily-memorable addresses.
The \hitlist occupies a middle ground of sorts, discovering both core
Internet infrastructure, as well as some higher-entropy addresses
assigned to CPE devices at the network periphery.
In the next subsection, we further investigate address entropies within
our \ds to illuminate differences in \vsix addressing schemes between
service providers.

\parhead{Observed address durations} %
The durations over which we observe distinct \vsix addresses vary
significantly, as shown in Figure~\ref{fig:lifetimes}.
Figure~\ref{fig:all-lifetimes} displays a CCDF of the lifetimes with
which we observe all of the 7.9 billion addresses in our \ds.
More than 60\% of them are observed only once (a ``lifetime'' of 0
seconds in the plot).
With purely passive measurements, we cannot determine whether this is
because the devices had highly ephemeral addresses, or simply because
they only contacted our (or anyone's) NTP servers only once.
This demonstrates the importance of combining passive collection with
active scans; so long as a device shows up even once at our servers
(e.g., at boot-up), it can be used in subsequent backscanning.

At the opposite extreme, 95,780,865 (1.2\%) \vsix addresses are
observed for a week or longer, 32,985,774 (0.4\%) \vsix addresses for a month or
longer, and 2,218,998 (0.03\%) \vsix addresses for more than six months.
Figure~\ref{fig:iid-lifetimes} is a CDF of the 670,737,407 unique IIDs
we observed, binned by the entropy of the IID.
This figure shows that while $\sim$10\% more of the low normalized
entropy ($<0.25$) IIDs appear only once in our corpus than medium or
high entropy IIDs, low entropy IIDs are more likely to persist for long
periods of time.
In fact, 10\% of all low entropy IIDs are observed for a week or more,
as compared to 5\% or less of the medium and high entropy IIDs.

\parhead{Summary} %
Collectively, these results show that our hitlist of IPv6 addresses
has little overlap with prior datasets, and in particular adds more
end-host devices than any prior efforts.
This is a natural progression in IPv6 hitlists, but one that would have
been much more difficult with purely active measurement-based
approaches.

\begin{figure}[t]
\centering
    \includegraphics[width=\linewidth]{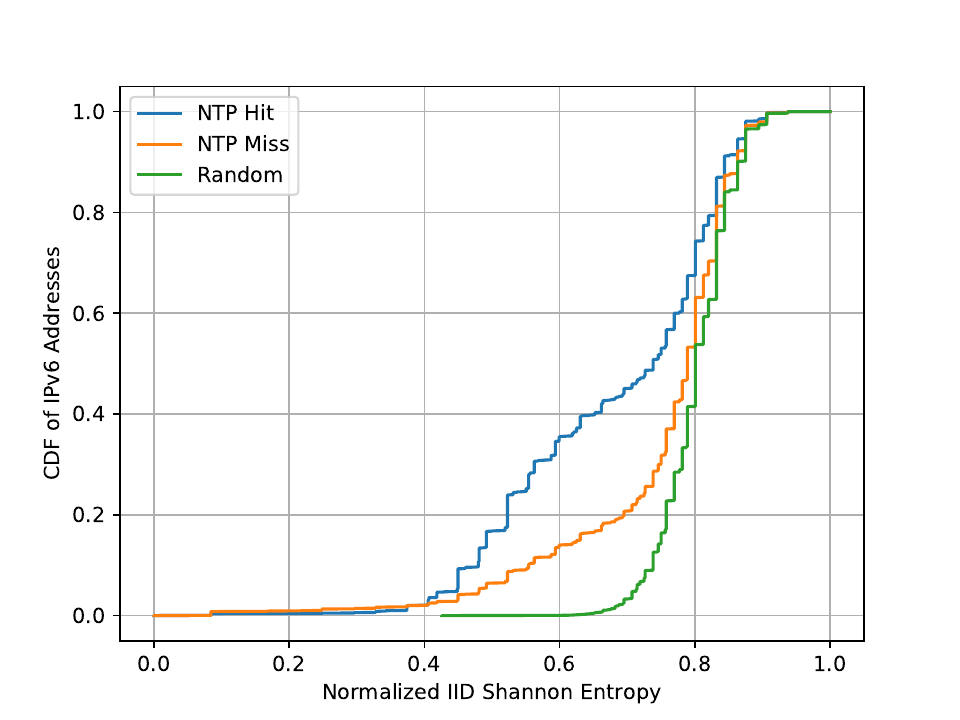}
    \caption{CDF of the IID entropy of NTP clients probed back with \yarrp and
        \zmapsix. The clients that we passively detect from running an NTP
	client are responsive to back scanning, particularly those with
	slightly lower entropy in their IID.}
\label{fig:backscan-entropy}
\end{figure}

\subsection{Backscanning and Aliased Networks} %

The IPv6 addresses we discovered are only useful for Internet scanning
insomuch as they are responsive to outside scans.
Moreover, if the addresses we learned are merely aliases of one
another, then the volume of them would not be useful.
To test both of these, we initiated \yarrp traces and \zmapsix probes from
five NTP servers back to the clients that contacted them, as well as to
a random address in the same /64 (see \S\ref{sec:methodology}).

\parhead{Responsiveness to backscanning} %
    As described in \S\ref{sec:methodology}, we initiated a limited number of
    \icmpsix scans back to NTP clients over the course of a week in January 2023. About two-thirds of the 71,341,581 NTP clients that were probed from
the NTP servers responded to \yarrp or \zmapsix probes.
This shows that the addresses we obtained can be used as scan targets.
In contrast to the NTP clients we probed, the random targets we probed
in the same network as the NTP clients responded only 3.5\% of the
time.
Because it is exceedingly unlikely that we guessed a live random-IID
\vsix address, these responses almost certainly originate from aliased
networks.

Figure~\ref{fig:backscan-entropy} displays the IID entropy of the
responsive addresses broken down by whether the address was responsive
to backscanning (``NTP hit'') or not (``NTP miss'') or if the address
was randomly chosen yet responsive (``Random'').
The responsive NTP client addresses exhibit the lowest median
normalized IID Shannon entropy, although they still exhibit higher
entropy than addresses in our comparison datasets.

The higher the IID entropy, the less likely the end-host was to
respond.
Nearly 70\% of the unresponsive NTP clients had normalized entropy greater than 0.75,
compared to only $\sim$50\% of the responsive clients.
We speculate that this can be attributed to infrastructure devices,
which typically have stable, low-entropy IIDs, and are more likely to
be responsive than client devices, such as mobile phones and personal
computers.
Client devices are more likely to reside behind a router or CPE device,
which are often configured to block unsolicited inbound traffic, such
as our \icmpsix \yarrp and \zmapsix probes.
Also, because \vsix client addresses are often 
ephemeral, it is conceivable that some clients change addresses after
querying our NTP servers and are no longer assigned that address when we probe
it
after the ten minute interval expires. Mobility may also play a factor, with
devices switching to a new network in the period between sending an NTP request
and when our probing began.

\parhead{Discovering aliased networks} %
As part of our backscanning, we received \icmpsix responses to 4,476,089 unique, random
\vsix addresses we probed. These responses originated from 3,740,619 unique /64s.
Because these addresses were chosen randomly from within active /64s, it is much
more likely that the network in question is aliased rather than we randomly
chose the address of a live host in an unaliased network.
We compared our inferred aliased networks to those within the IPv6
Hitlist, which maintains a list of aliased networks.
Of the 4.5 million aliased addresses we probed, the \hitlist also
categorized 4,425,001 (98\%) as aliased.
However, we discover an additional 46,512 aliased addresses in prefixes
the \hitlist does not recognize as aliased.
This suggests that backscanning NTP clients is a potential avenue for
discovering additional aliased \vsix networks.

\begin{figure*}[t]
    \begin{subfigure}[c]{.48\textwidth}
    \centering
    \includegraphics[width=\linewidth]{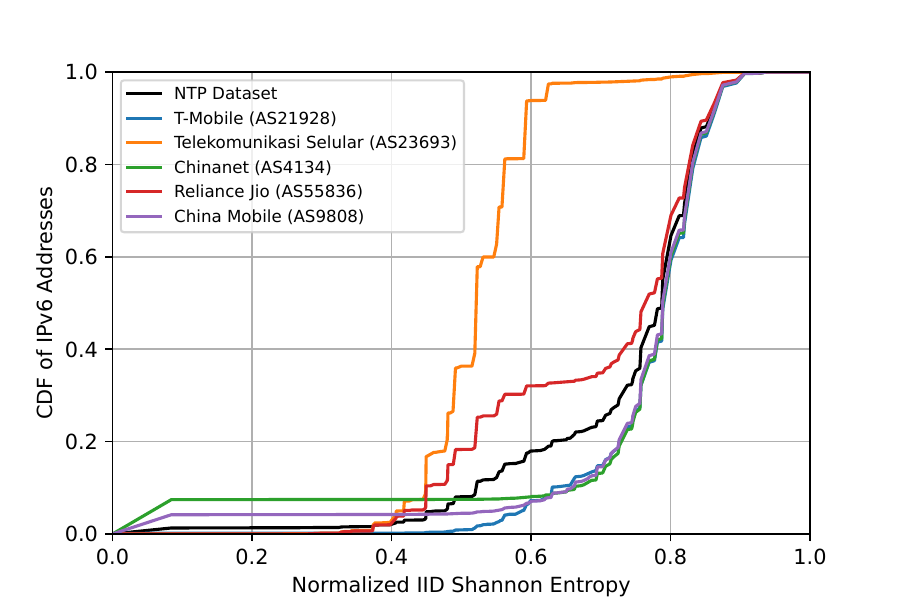}
    \caption{CDF of \vsix address IID entropies observed in the top five ASes
        observed between Jan-Aug 2022.}
        \label{fig:entropy-top-5-dataset}
    \end{subfigure}%
	\hfill
    \begin{subfigure}[c]{.48\textwidth}
    \centering
    \includegraphics[width=\linewidth]{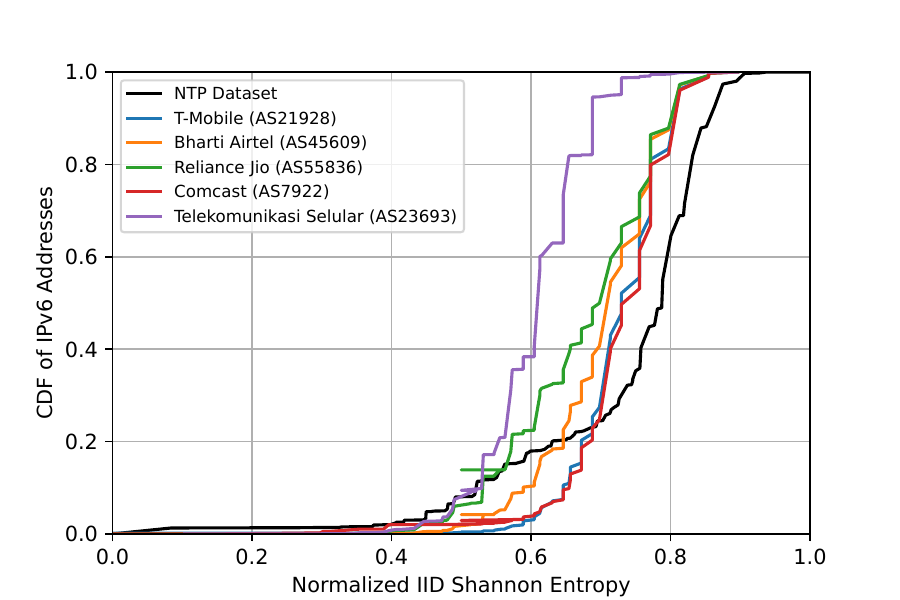}
    \caption{CDF of \vsix address IID entropies for the top five ASes
        observed on 1 July 2022.}
    \label{fig:entropy-top-5-day}
    \end{subfigure}
    \caption{A comparison of the normalized Shannon entropies of collected \vsix
    addresses over two time periods.}
    \label{fig:entropy}
\end{figure*}

Finally, we examine the NTP clients in networks that we later
determined were aliased via our backscanning technique.
We find 3,841,751 NTP client addresses are part of aliased /64s as
discovered by backscanning.
These NTP client addresses originate from 36 different ASes.

We note that because the prefixes these NTP clients originate in are
aliased, an active measurement campaign would be unable to distinguish
these live hosts from aliased responses.
That is, if it even attempted to scan the prefix in the first
place---filtering known aliased networks is a best practice first step
when conducting active measurements.
Indeed, we searched for these addresses in a contemporaneous \vsix
Hitlist and found \emph{only 23} addresses in our backscanning-detected
/64s, while our NTP corpus contains 3,841,751.

\parhead{Summary} %
These results show that the majority of the addresses learned from our
NTP-based dataset are responsive to scanning, despite the fact that
most of them are clients and thus likely behind CPE devices.
While this is a boon for Internet scanning, the fact that many of the
devices are clients means that it is also a potential security issue;
most would likely benefit from being behind firewalls.
We also find that many of the addresses in our dataset are unlikely to
ever be discovered with today's active measurement techniques: guessing
active random \vsix addresses or differentiating active addresses in aliased
networks is
impractical.
Collectively, these results motivate complementing active measurements
with passive data collection to obtain more complete IPv6 hitlists.

\subsection{\vsix Addressing Patterns} %

Our collection of \vsix addresses permits insight into address allocation
patterns, both on a macro scale, which has been previously
studied~\cite{Foremski:2016:EUS:2987443.2987445,expanse},
and also at an AS level.
The magnitude of addresses we
obtain and the breadth of the networks from which they originate allows us to
observe interesting phenomena specific to individual service providers.
These
phenomena are not visible from active measurements, and may help inform
active measurement studies by illuminating the types of addresses present in
various ASes.

Figure~\ref{fig:entropy} displays two plots; each is a CDF over the
total number of \vsix addresses observed during a time interval plotted
against the normalized Shannon entropy of address \ac{IID}. %
While an imperfect proxy for randomness (\eg the \ac{IID}
\texttt{0123:4567:89ab:cdef} has a normalized Shannon entropy of 1.0,
but such a clear pattern might conceivably appear as the result of a
network operator manually assigning it), \ac{IID} entropy allows us to
make generalizations about the types of addresses found within specific
\acp{AS}.

\begin{table*}[htb!]
    \small
\begin{tabular}{lrlr}
  \textbf{Manufacturer} & \textbf{Count}     & \textbf{Manufacturer} & \textbf{Count}   \\\hline
Unlisted & 126,789,603 & Sunnovo International Limited & 1,193,746 \\
Amazon Technologies Inc. & 19,090,527  & Hui Zhou Gaoshengda Technology Co.,LTD
    & 1,067,459 \\
Samsung Electronics Co.,Ltd & 2,683,846   & Huawei Technologies
    & 876,083  \\
Sonos, Inc. & 1,633,209   & Shenzhen Chuangwei-RGB Electronics & 861,122  \\
vivo Mobile Communication Co., Ltd. & 1,330,987   & Skyworth Digital Technology
    (Shenzhen) Co.,Ltd & 723,316\\
\end{tabular}

	\vspace{.1in}
    \caption{Number of MAC addresses extracted from EUI-64 IPv6 NTP clients by
    manufacturer ($N$ = 171,611,786)}
    \label{tab:manuf_counts}
\end{table*}

\parhead{Variability in entropy across ASes}
Figure~\ref{fig:entropy-top-5-dataset} displays CDFs of the normalized \ac{IID}
Shannon entropies of the top five most commonly-observed \acp{AS} in
our entire dataset, collected between January and August 2022.
Three of the top five ASes (T-Mobile, ChinaNet, and China Mobile)
exhibit entropy behavior that closely tracks with the NTP aggregate
curve shown in Figure~\ref{fig:entropy-compare}.
The remaining two entropy curves, representing Reliance Jio and
Telekomunikasi Selular (an Indian and Indonesian mobile provider,
respectively), exhibit much lower median entropies.
Further, while about 60\% of the Reliance Jio addresses have high
entropy ($>$0.75), approximately one-third exhibit 
a lower entropy below 0.6.
This indicates that there are perhaps multiple classes of \vsix
addresses reaching our NTP servers from this network.
Closer inspection reveals that at least two addressing patterns exist
for hosts on Reliance Jio's network: one that randomizes all eight bytes of the
IID, and another that uses the only lower four IID bytes with the remaining four set to 0.

\parhead{Addressing strategies}
This result inspired us to investigate different patterns of addresses
within ASes.
To permit comparison to IPv6 Hitlist data, which is released in snapshots at
intervals, 
 we limit this analysis to a
single day: 1 July 2022. Limiting this analysis to a single day also minimizes
the influence that numerous random, ephemeral addresses from the same host might
have over longer observation windows.
Figure~\ref{fig:entropy-top-5-day} shows the entropy of the top five ASes
from that day in our dataset.

We compare the 46,195,900 NTP addresses we collect on 1 July 2022 to
the \hitlist's 2,970,366 IPs released on 1 July for the week prior
across seven categories:
(1)~All zero IIDs (``Zeroes'');
(2)~IIDs with only the least significant byte set (``Low Byte'');
(3)~two least significant bytes (``Low 2 Bytes'');
(4)~IPv4 mapped addresses; and 
(5)~high-entropy ($>$0.75);
(6)~medium-entropy (between 0.25 and 0.75); and
(7)~low-entropy ($<$0.25) IIDs.
For IPv4 mapped addresses, which embed an IPv4 address in the \vsix
IID, we check whether any of three different embedded address encodings produce
an \vfour address
in the same AS as the \vsix address they are embedded in.
We accept IPv4 embedded addresses only when i) there are at least 100
instances of them in the AS, and ii) more than 10\% of the AS's total
addresses are \vfour embedded. These steps reduce false positives from
random IIDs that coincidentally produce embedded IPv4 addresses in the same AS
as the \vsix address.

\begin{figure}[t]
\centering
    \includegraphics[width=\linewidth]{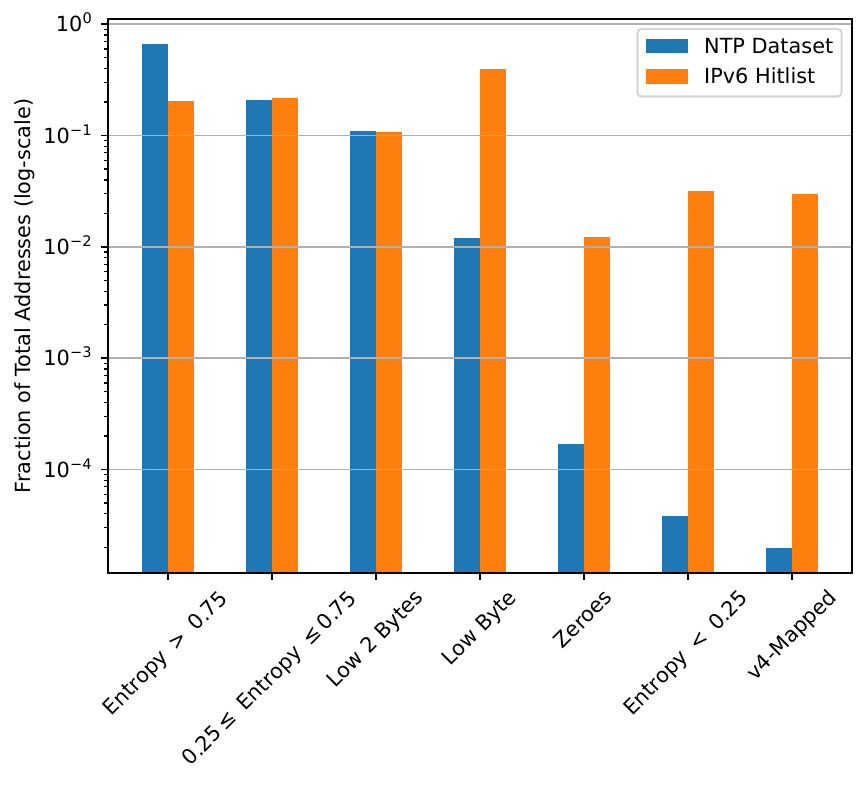}
    \caption{Fraction of the NTP corpus and \hitlist that fall into each of
    seven categories for 1 July 2022. Note: $y$-axis is log-scale.}
\label{fig:type-compare}
\end{figure}

Figure~\ref{fig:type-compare} compares the frequency of these seven
categories between our NTP-derived dataset and the IPv6 Hitlist.
We find that the distributions of address types vary significantly
between the two datasets. For the day considered, the NTP dataset is
two-thirds high-entropy, with an additional 21\% medium entropy. The
\hitlist, on the other hand, is approximately 20\% medium and high
entropy over the same time period.  The fraction of \hitlist Low Byte
addresses, however, is nearly 33 times that of the NTP corpus, and it
contains 3\% \vfour mapped addresses compared to the NTP corpus' 0.00002\%. 

Taking only the \hitlist into consideration,
it would appear that the preponderance of IPv6 addresses' IIDs have only the
least significant bytes set (``Low Byte'').
But in our NTP dataset, the majority of active IIDs
are in fact high-entropy.
We believe that the Low Byte are more likely to be routing
infrastructure; it is far easier for a network operator to manually
set, read (e.g., in logs), and remember an IID with few bytes set
(e.g., \texttt{::100}) than random ones.
Indeed, because the IPv6 Hitlist relies heavily on \texttt{traceroute}-like
techniques,
we suspect it comprises many such routers.
Nevertheless, the reality of the IPv6 Internet is that the majority of addresses
are randomly set by clients: precisely the kinds of addresses that our
prior results show are impractical for active measurements to obtain at
scale.

\parhead{Summary}
This example application of our NTP-derived dataset shows that
\emph{larger, more client-rich hitlists stand to improve future network
measurement studies.}
The push for larger hitlists is justified, at least from a measurement
perspective.
In the next section, we turn to whether larger hitlists come at
increased security cost.

 %


\section{Privacy Issues of Larger Hitlists} 
\label{sec:security}

\begin{figure*}[t]
    \begin{subfigure}[c]{.48\textwidth}
\centering
    \includegraphics[width=\linewidth]{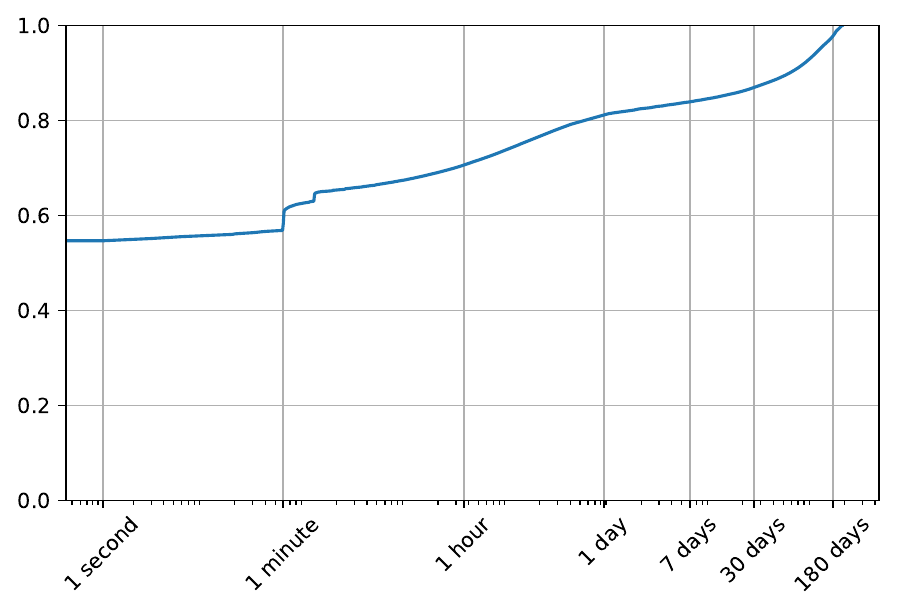}
    \caption{Lifetime of observed EUI-64 IIDs during the seven months of our
    study ($x$-axis logscale).}
\label{fig:eui-lifetime}
\end{subfigure}
\hfill
\begin{subfigure}[c]{.48\textwidth}
\centering
    \includegraphics[width=\linewidth]{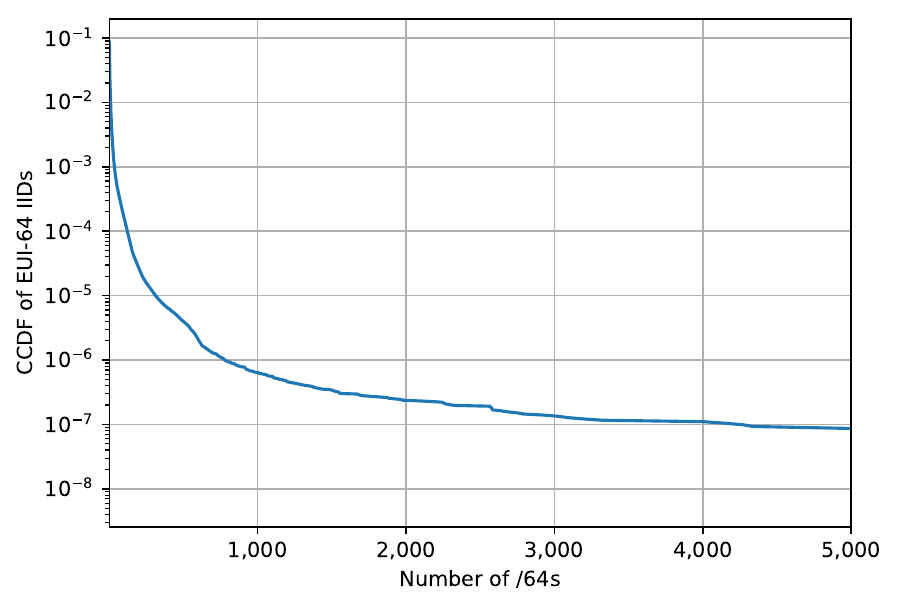}
    \caption{CCDF of EUI-64 IIDs depicting the number of /64s each EUI-64 IID
    appears in.}
\label{fig:eui-prefixes}
\end{subfigure}
    \caption{EUI-64 IIDs permit long-term tracking of devices as they
    transition between network prefixes.}
    \label{fig:eui}
\end{figure*}

In this section, we perform what is, to our knowledge, the first
empirical analysis of the privacy leakages in large, public hitlist
datasets.
We study two sources of privacy leakage: tracking via \eui, and
geolocation by matching MAC addresses to geolocated BSSIDs (Basic Service Set
Identifier, the MAC address of
a WiFi access point), as
in Rye and Beverly~\cite{ipvseeyou}.
Our analysis here shows the potential harms inherent in larger
hitlists.

\subsection{Prevalence of \eui \acp{IID}} %

Both of the techniques we use to track and geolocate users make use of
\eui \acp{IID}.
EUI-64 \vsix addresses have long been considered
privacy vulnerability and have recently been studied extensively in CPE
devices~\cite{rye2021follow,ipvseeyou} and in traffic from an
ISP~\cite{gasserapple}. As our corpus consists primarily of client devices
(\S\ref{sec:measurements}) from a global network of NTP server vantage points,
we were optimistic that \eui \vsix address prevalence would be low. Unfortunately, this was
not the case.

Our \ds contains 238,281,703 \eui \vsix addresses: 3\% of our corpus 
and more than the total
number of \emph{all} \vsix addresses reported in our comparison
datasets (see Table~\ref{tab:datasets}).
Moreover, we 
can be certain that these are not randomly-generated addresses that
\emph{appear} to be \eui due to \texttt{0xFF 0xFE} in the fourth
and fifth bytes of the \ac{IID}.
The probability that a randomly-generated IID matches those bytes is $2^{-16}$.
Therefore,
we would expect $\frac{7,914,066,999}{65,536}$ randomly-generated apparent \eui \vsix
addresses, which is less than
121,000.

Among the \eui \vsix addresses present in our corpus, we find 171,611,786 unique embedded MAC addresses.
MAC addresses can appear in multiple different \vsix addresses for
several reasons.
First, when a device that uses \eui addresses changes networks (\eg due to
mobility or a prefix rotation), its address will change but its
\eui will remain the same.
Second, some device manufacturers have reused MAC address space, or
assign common patterns (\eg \texttt{00:00:00:00:00:00}), which then
appear in \eui \vsix addresses for multiple devices.

To better understand the types of devices using \eui addresses in our
corpus, we first resolve the OUI from the MAC address extracted from
the \eui \ac{IID} to the manufacturer listed in the IEEE's OUI
database. We do this by removing the \texttt{0xFF 0xFE} from bytes four and five
of the IID, and then inverting the Universal/Local bit of the resulting MAC
address (the second-least significant bit of the first byte) if it is set.
Table~\ref{tab:manuf_counts} contains the 10 most frequently observed
manufacturers.

Surprisingly, the most common OUI we observe (126,789,603 MAC
addresses, or 73.9\% of all observed MACs in our dataset) is
``Unlisted''---that is, they could not be resolved to a manufacturer at all.
These are not merely random addresses; recall that we would expect
fewer than 121,000 random IIDs to look like \eui.
Moreover, manually inspecting the ``Unlisted'' addresses, we see significant  
MAC address counts in OUIs that are not in the IEEE OUI database.
For instance, the most common ``Unlisted'' OUI is \texttt{F0:02:20},
with 52,218 distinct MACs embedded in \eui addresses.
The frequency with which this OUI appears in \eui addresses makes it
unlikely that these MAC addresses appear through a random process.
On the other hand, 42,901 OUIs we classify as ``Unlisted'' appear in
only one MAC address embedded in an \eui address; we believe these are
the randomly generated IIDs that appear to be \eui.

Included among the other nine most common manufacturers 
are makers of popular mobile, smart home, and IoT devices.

\begin{figure*}[t]
\begin{subfigure}[c]{.49\textwidth}
\centering
    \includegraphics[width=\linewidth]{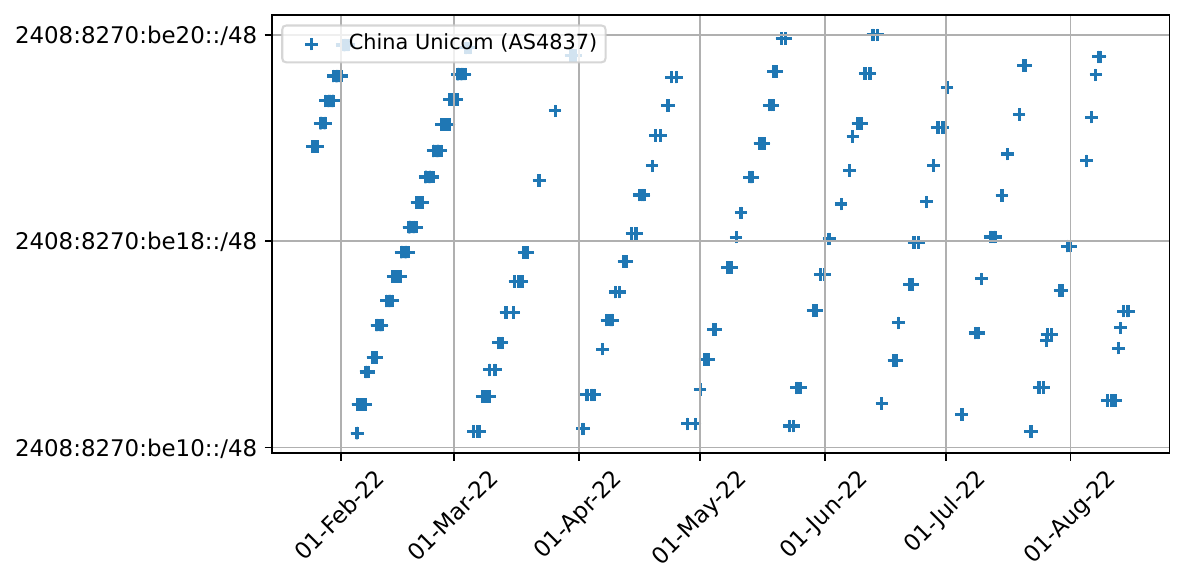}
    \caption{A MAC from an unregistered OUI (\texttt{A8:AA:20}) is frequently
    renumbered within the same AS.}
\label{fig:renumbering}
\end{subfigure}
\hfill
    \begin{subfigure}[c]{.49\textwidth}
\centering
    \includegraphics[width=\linewidth]{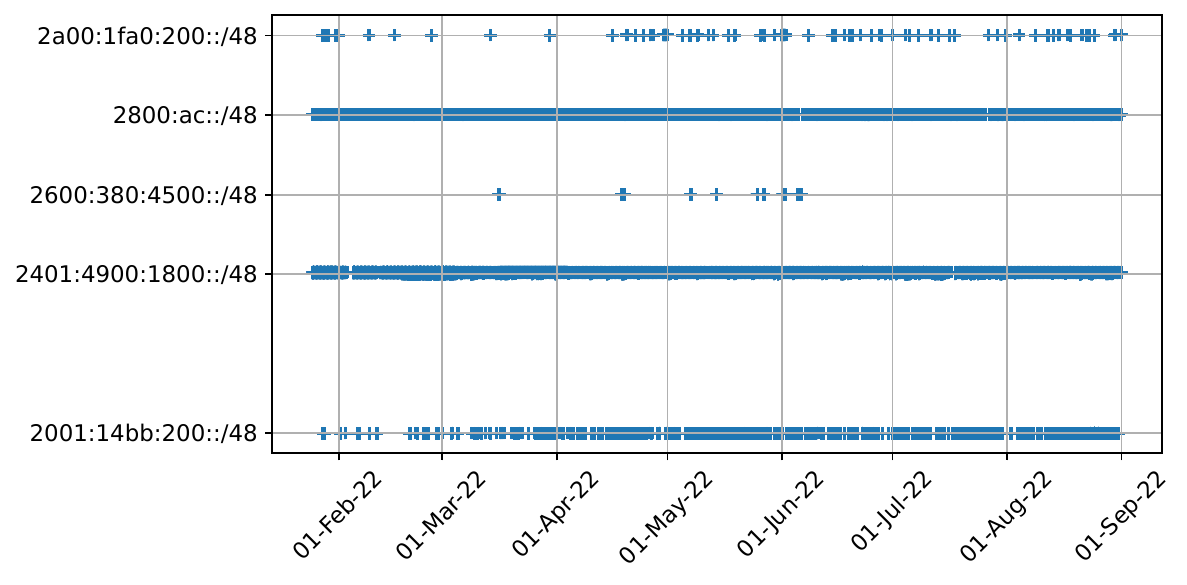}
    \caption{A single MAC address appears in EUI-64 IPv6 addresses in 70 different ASes.}
\label{fig:reuse}
\end{subfigure}
\hfill
\begin{subfigure}[c]{.49\textwidth}
\centering
    \includegraphics[width=\linewidth]{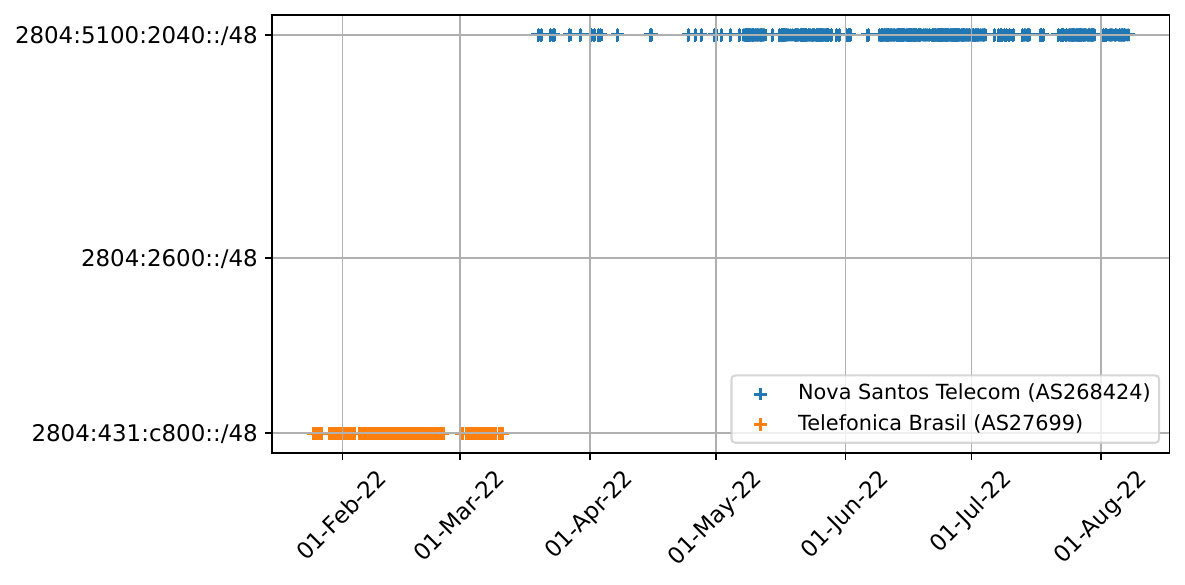}
    \caption{A device changing service between two Brazilian providers.}
\label{fig:change-provider}
\end{subfigure}
\hfill
\begin{subfigure}[c]{.49\textwidth}
\centering
    \includegraphics[width=\linewidth]{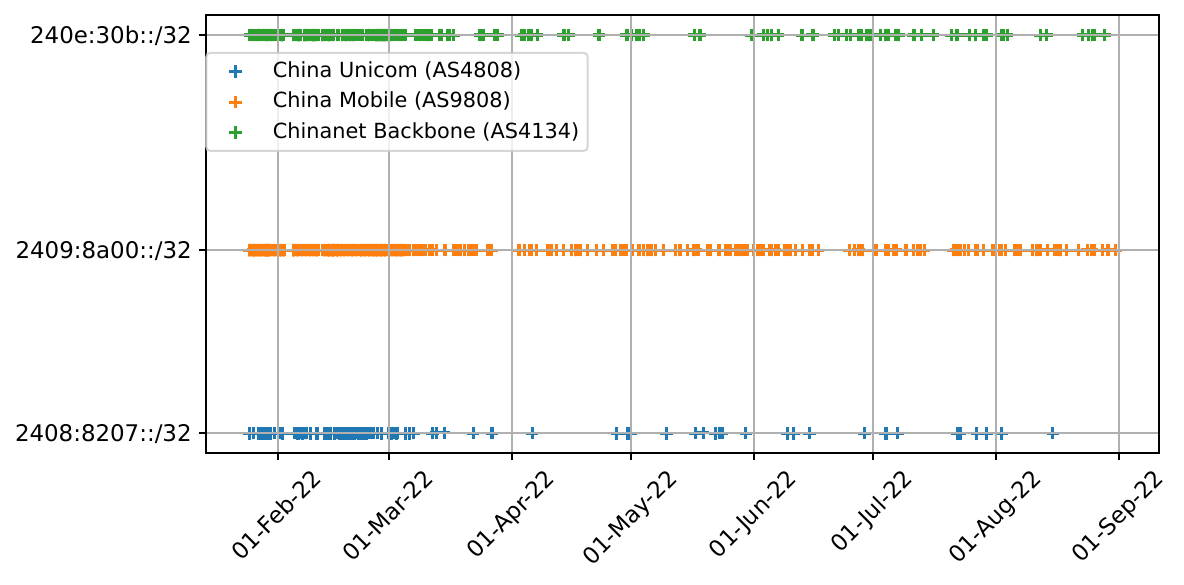}
    \caption{A Huawei MAC address frequently moves between multiple Chinese
    ASes.} 
    \label{fig:pol-example}
\end{subfigure}
    \caption{A variety of \eui IID tracking situations arise through the use of
    \eui \vsix addresses.}
    \label{fig:tracking-types}
\end{figure*}

\subsection{Tracking \eui \acp{IID}} %

Here, we evaluate the extent to which the \eui IIDs in our dataset
could be potentially used for tracking users.

Figure~\ref{fig:eui-lifetime} depicts a CDF of the lifetime of all \eui
IIDs over the seven months of our study.
While this generally tracks with all IIDs (see
Figure~\ref{fig:iid-lifetimes}) \eui IIDs  are less likely to be
observed only once ($\sim$55\% compared to 60--70\%) and exhibit the
same long, fat tail that low-entropy IIDs do.
This is due to the fact that when devices using \eui change networks,
they continue to use the same \eui IID.

The lengthy observation window we observe for many \eui{} IIDs demonstrates
that \emph{a passive adversary can leverage a large, longitudinal
hitlist to track an unsuspecting user's device.}
Because many service providers frequently ``rotate'' prefixes delegated
to customers, often on timescales on the order of days or weeks, the
ability to track a device by \eui passively offers a major advantage
over active techniques.

Figure~\ref{fig:eui-prefixes} displays a CCDF of the number of /64s
each \eui in our corpus appears in.
While most \eui IIDs appear in only one /64 prefix, many appear in
dozens, hundreds, or even thousands of /64s during the seven months of
our study.

To determine which of the \eui IIDs could be trackable users, we apply
the following heuristics-based approach.
We compute, for each \eui IID:
(1)~The number of ASes it appears in; if more than 1, then we call
it ``high,'' otherwise ``low.''
(2)~The number of countries it appears in; if more than 1, then we call
it ``high,'' otherwise ``low.''
(3)~The number of transitions between different /64s it makes; if more
than 10, then we call it ``high,'' otherwise ``low.''
If a device never changes its /64, then we consider it not trackable;
of the 171,611,786 \eui MAC addresses we observed, 14,943,429 (8.7\%)
of them appear in at least two /64s.

Using this heuristic, we classify \eui IIDs into five categories as to
the likely explanation for their re-occurrence:

\parhead{Mostly static hosts}
The most common classification (12,853,055 of 14,943,429, or 86\%) is
\eui IID that are labelled ``low'' across all three categories.
These IIDs stay within the same AS and country throughout our
observation period of them, and if they change /64s, they do so
relatively rarely.

\parhead{Likely prefix reassignment}
The second-most common classification (1,215,400 or 8\%) appear in only
one AS and one country, but /64 transitions is ``high.''
One potential cause of this is service providers periodically
reassigning new delegated
prefixes to their customers
Because this behavior is often based on provider policy, we observe it
occurring more frequently in some providers than others.
Figure~\ref{fig:renumbering} displays an exemplar of this behavior.

\parhead{Likely MAC reuse}
In some instances (2,320 or 0.01\%), we detect a single \eui IID in a
large number of ASes and countries, accompanied by a high number of
transitions between /64.
In these cases, we believe that  are observing instances of MAC address
reuse by a manufacturer, and that we are detecting several devices
within different networks simultaneously.
Figure~\ref{fig:reuse} depicts a MAC address from \eui IIDs that
appear in a ``high'' number of countries and ASes.

\parhead{Changing providers}
We observe 5\% of devices in multiple ASes within the same country that
transition a ``low'' number of times between /64s.
This behavior could arise from a static IoT or CPE device changing
service providers.
Figure~\ref{fig:change-provider} displays an example of an \eui IID
seen transitioning between one AS to another. For the first month and a half of
our study, the device appeared only in Telefonica Brasil's network. Mid-March
and later, however,
the device appeared only in Nova Santos Telecom.

\parhead{Likely user movement}
Finally, 66,187 (0.44\%) \eui IIDs are observed in a high number of
ASes within the same country that are also classified as ``high'' in
/64 transitions.
One possible cause of this behavior is a mobile device using \eui
addressing; as it transitions from a home WiFi network to a cellular
network, it appears in multiple ASes and frequently changes /64s.
Figure~\ref{fig:pol-example} depicts a Huawei MAC address that moves
between three Chinese networks frequently over time.
We believe that these are indicative of addresses that permit user
tracking over time, and thus pose a risk to users' PII.
While low in percentage, the raw number of potential user-tracking
events this permits is large and concerning.

\subsection{Geolocation} %
\label{sec:geo}

Recently, Rye and Beverly~\cite{ipvseeyou} described a technique for
geolocating CPE routers by linking MAC addresses seen in \eui IIDs with
BSSIDs in publicly available wardriving datasets.
Here, we apply their technique to all of our \eui addresses.
Whereas their initial analysis relied on active scanning and thus
comprised mostly CPE routers, our dataset allows us to also consider
IoT and mobile devices within customers' LANs (see
Table~\ref{tab:manuf_counts}).

We query geolocation databases (e.g. WiGLE~\cite{wigle} and Apple and Google's WiFi Location
APIs~\cite{applegeo,googlegeo}) for WiFi BSSIDs in the same OUIs as the
171,611,786 MAC addresses we derive from \eui IIDs. This produces 
2,692,307 distinct WiFi BSSIDs with associated geolocation data.
Next, we use our \eui MAC addresses and wireless geolocation data to
infer the offsets between wired and wireless MAC addresses in the same
OUI.
For each MAC address embedded in an \eui IID, we compare it to each
wireless BSSID in the same OUI from our dataset, recording the offsets
between each pair of two identifiers.
We then tally the most common positive and negative offsets from the
wired MAC to a geolocated BSSID, and select the offset with the largest
number of MAC-to-BSSID matches as the ``correct'' offset.
In this manner, we generate wired-to-wireless offsets for 117 OUIs with
at least 500 wired MAC-to-BSSID pairs.
Finally, we used these offsets to determine how many of the \eui MAC
addresses we could match with WiFi BSSIDs and therefore geolocate.

All together, our methodology links 225,354 unique
MAC addresses from our collected NTP dataset with geolocated WiFi BSSIDs.
Although we do not have ground truth for the geolocations of these \eui \vsix
addresses, we note that prior work validated the efficacy of this technique with a large US
residential ISP~\cite{ipvseeyou}.

Our geolocations resolve to 140 different countries. However, a large majority
(174,155 or 75\%) of the geolocated \eui \vsix addresses are from Germany.
Mexico (7\%), India (4\%), France (3\%), and Luxembourg (2\%) round out the top
five countries of the geolocated \eui \vsix addresses. The over-representation
of Germany and neighboring countries is due to the preponderance of AVM GmbH,
the maker of the popular Fritz!Box router, MAC addresses in our geolocated \eui
\vsix address set. AVM MAC addresses are responsible for 180,727 (80\%) of the
geolocated \eui \vsix addresses. Prior communication with AVM product security
personnel confirmed this geolocation vector exists, and Fritz!OS version 7.50 eliminated
\eui WAN addresses support in December 2022.

Due to the prevalence of client addresses in our passive NTP corpus, using Rye
and Beverly's CPE router geolocation technique permits fewer \eui \vsix
geolocations than did their original study~\cite{ipvseeyou}. That work
specifically targeted CPE with active measurements. However, we demonstrate that
our NTP dataset contains some number of CPE routers that visit the NTP Pool for
time that are susceptible to this privacy attack. Further, it is entirely
passive. The only defense from this form of geolocation and tracking is to
sever the linkage between the MAC addresses that appear in an \eui IPv6
address and the BSSIDs that the WiFi access points use. Due to the potential for
device tracking detailed earlier in this section, we recommend the use of random
\vsix addresses.

%

 %

\section{Conclusions} 
\label{sec:conclusions}

In this work, we accumulated the largest hitlist of \vsix addresses solely
through publicly-available means, without the aid of a CDN or ISP.  
We collected
7.9 billion unique addresses from a distributed set of 27 NTP servers located in
cloud providers around the world. In addition to containing orders of
magnitude more live addresses than existing hitlists, our dataset differs from
current lists in that it contains \emph{different types of addresses.} The
addresses we obtain are highly entropic and ephemeral. They often come from
client devices, as the OUIs of the embedded MAC addresses in \eui \vsix
addresses show. And, crucially, these addresses are almost entirely absent in
state-of-the-art hitlists today, which biases these hitlists toward addresses
that can easily be discovered with active measurements or the DNS, like
infrastructure devices and servers. 

The ability to capture massive numbers of active client devices raises ethical
questions not previously raised by active measurements. Since many of these
addresses are highly random and ephemeral, they are likely uniquely associated
with a single device or individual at a specific moment in time. Because of this
uniqueness, we believe that these addresses deserve a special level of care in
handling. As such, we will release our dataset truncated to the /48 level.

Finally, we repeat a plea for manufacturers to discontinue the use of \eui \vsix
addresses.  Our results show that \eui addressing is not uncommon among NTP
clients, and is implemented by popular manufacturers of IoT and smart home
products. The use of these addresses permits, in some cases, tracking of devices
across networks, as well as fine-grained geolocation through correlation with
wireless identifiers.
 
%


\section*{Acknowledgments} 

We thank Ask Bj{\o}rn Hansen and Steven Sommars from the NTP Pool project
for their support and feedback on data sharing considerations.
We also thank Rob Beverly and Justin Rohrer for early feedback and
infrastructure support, as well as the anonymous reviewers for their helpful
comments.
This work was supported in part by NSF grants CNS-1901325 and
CNS-1943240.


\bibliographystyle{ACM-Reference-Format}
\bibliography{conferences,refs}

\begin{acronym}
  \acro{AS}{Autonomous System}
  \acrodefplural{AS}[ASes]{Autonomous Systems}
  \acro{ASN}{\ac{AS} Number}
  \acro{BGP}{Border Gateway Protocol}
  \acro{CDN}{Content Distribution Network}
  \acro{CPE}{Customer Premises Equipment}
  \acro{DAD}{Duplicate Address Detection}
  \acro{EUI}{Extended Unique Identifier}
  \acro{ISP}{Internet Service Provider}
  \acro{IID}{Interface Identifier}
  \acro{LAN}{Local Area Network}
  \acro{NIC}{Network Interface Card}
  \acro{NTP}{Network Time Protocol}
  \acro{MAC}{Media Access Control}
  \acro{OS}{Operating System}
  \acro{OUI}{Organizationally Unique Identifier}
  \acro{SOHO}{Small Office-Home Office}
  \acro{U/L}{Universal/Local}
  \acro{SLAAC}{Stateless Address Autoconfiguration}
  \acro{VPS}{Virtual Private Server}
\end{acronym}

\end{document}